\documentclass[trackchanges]{aastex631}
\usepackage{bm}
\usepackage{amsmath}
\usepackage{url}
\usepackage{natbib}
\usepackage{xcolor}

\begin{document}
\title{Preferential Positron Acceleration in Relativistic Magnetized Electron-Positron-Ion Shocks}

\correspondingauthor{Shori Arai}

\author{Shori Arai}
\affiliation{Graduate School of Science and Engineering, Chiba University, 1-33 Yayoi, Inage-ku, Chiba, Chiba 263-8522, Japan}
\email{24wm2101@student.gs.chiba-u.jp}

\author[0000-0002-1484-7056]{Yosuke Matsumoto}
\affiliation{Institute for Advanced Academic Research, Chiba University, 1-33 Yayoi, Inage-ku, Chiba, Chiba 263-8522, Japan}
\email{ymatumot@chiba-u.jp}

\begin{abstract}
Relativistic shocks are considered efficient accelerators of charged particles and play crucial roles in high-energy astrophysical phenomena, such as gamma-ray bursts and pulsar winds. This study focuses on positron accelerations in magnetized relativistic shocks in electron-positron-ion plasma. Employing one-dimensional ab initio particle-in-cell simulations, we found a preferential positron acceleration through an interaction with the wakefield associated with a precursor wave in the upstream region. Test particle simulations revealed that the selective acceleration occurs for sufficiently large amplitudes of the wakefield. The mechanism can be understood as the relativistic $\boldsymbol{E}\times\boldsymbol{B}$ acceleration formulated in the upstream frame. A theoretical analysis of the positron acceleration in astrophysical contexts is presented, supporting ultra-relativistic shocks in pulsar winds as a primary source for the high-energy positron excess.
\end{abstract}


\section{Introduction} \label{sec:Intro}
Cosmic rays have been studied extensively since their discovery over a century ago. However, many unsolved issues remain. One of the open questions is the origin of the high-energy positron cosmic rays. Antiparticles are generated through collisions between primary cosmic ray protons and the interstellar medium, resulting in secondary cosmic rays. The energy spectrum of secondary positrons can be understood from the pion creation rate during these collisions, the subsequent diffusion and transport of generated particles within the galactic magnetic fields, and their energy losses due to synchrotron radiation and the inverse Compton scattering \citep[e.g.,][]{moskalenkoProductionPropagationCosmicRay1998,strongCosmicRayPropagationInteractions2007}. However, space experiments such as the Payload for Antimatter Matter Exploration and Light-nuclei Astrophysics (PAMELA) \citep{adrianiCosmicRayElectronFlux2011,adrianiCosmicRayPositronEnergy2013}, the Fermi Large Area Telescope (Fermi-LAT) \citep{ackermannMeasurementSeparateCosmicRay2012}, and the Alpha Magnetic Spectrometer on the International Space Station (AMS-02) \citep{aguilarElectronPositronFluxes2014} have reported that the flux ratio of positrons to electrons, the positron fraction, at energies above 10 giga electron volts (GeV) exceeded expectations from the theories that are based on the secondary positron model, implying existence of other sources of high-energy primary positrons. Their origin has remained a topic of debate \citep[e.g.,][]{jinAstrophysicalBackgroundDark2020,evoliGalacticFactoriesCosmicray2021}. Among possible scenarios, we focus on relativistic collisionless shocks as the source of the high-energy primary positrons.

Collisionless relativistic shocks have been considered efficient particle accelerators and the primary sources of high-energy cosmic rays. Detailed mechanisms of charged particles' accelerations have been studied using the ab initio particle-in-cell (PIC) simulations. In unmagnetized or weakly magnetized shocks, electrons and positrons are scattered by turbulent magnetic fields generated by the Weibel instability, and accelerated through the first-order Fermi mechanism \citep[e.g.,][]{spitkovskyParticleAccelerationRelativistic2008,sironiMAXIMUMENERGYACCELERATED2013}. In mildly magnetized cases, the electron (positron) synchrotron maser instability at the shock front plays a crucial role as it generates precursor electromagnetic waves that propagate upstream \citep[e.g.,][]{langdonStructureRelativisticMagnetosonic1988,gallantRelativisticPerpendicularShocks1992,plotnikovSynchrotronMaserEmission2019}. The precursor waves are particularly important in ion-electron shocks, since they generate the electrostatic field, known as the wakefield, in the upstream region, leading to electron accelerations similar to those known as the plasma wakefield acceleration \citep[e.g.,][]{lyubarskyElectronIonCouplingUpstream2006,hoshinoWakefieldAccelerationRadiation2008,iwamotoPrecursorWaveAmplification2019}. 

While positron accelerations in pair plasma shocks have been extensively studied, investigations on positron accelerations in the presence of ions have been limited. In magnetized shocks, positrons could be preferentially accelerated by absorbing the high-harmonic ion cyclotron waves emitted by the ions in the downstream region \citep{hoshinoRelativisticMagnetosonicShock1992,amatoHeatingNonthermalParticle2006,stockemACCELERATIONPERPENDICULARRELATIVISTIC2012}. The acceleration proceeds more efficiently as the positron fractions to the electrons decrease. In unmagnetized or weakly magnetized cases, however, the energy spectra of positrons are much softer than electrons, regardless of the positron fraction, due to the deceleration by a longitudinal electric field ahead of the shock \citep{groseljMicrophysicsRelativisticCollisionless2022}. While the former results suggest an application to radio emissions in the pulsar winds, the latter results have been integrated into the gamma-ray bursts afterglow model.

In this study, we investigate positron accelerations in relativistic magnetized electron-positron-ion shocks accompanying precursor waves in the upstream region. We demonstrate a preferential positron acceleration through an interaction with the wakefield by employing one-dimensional (1D) PIC simulations and discuss its implications for the high-energy primary positrons suggested by the space experiments. The paper is organized as follows. First, we describe our simulation setup in Section \ref{sec:Setup}. Next, we describe the overall structures of relativistic magnetized shocks in Section \ref{sec:PIC}. Section \ref{sec:TestParticle} shows results from test particle simulations to explain the acceleration mechanism of positrons and electrons during the interaction with the wakefield. In Section \ref{sec:accel_conditions}, we derive conditions for the efficient acceleration based on the theory of the relativistic $\boldsymbol{E} \times \boldsymbol{B}$ motion and the precursor wave emission efficiency, and discuss possible applications to astrophysical objects. Finally, Section \ref{sec:Discussion} summarizes the results and discusses future perspectives.

\section{Simulation Setup} \label{sec:Setup}
We examine relativistic magnetized shocks in electron-positron-ion plasma using a 1D version of the Wuming PIC code \citep[][]{matsumotoWumingPIC2025}. We adopt Vay's particle integration algorithm \citep{vaySimulationBeamsPlasmas2008} to track particles' motions accurately in relativistic magnetized shocks. The simulation configuration involves the continuous injection of a cold plasma flow from the right-hand boundary of the computational domain in the $-x$-direction with the magnetic field imposed in the $+z$-direction. Particles undergo specular reflection at the left-hand boundary, where a conducting wall boundary condition is applied for the electromagnetic field. The interaction between the reflected particles and the incoming plasma facilitates the formation of a shockwave propagating in the $+x$-direction, establishing the simulation frame as the downstream rest frame. All physical quantities for space and time are in units of the electron inertial length $c/\omega_{\rm{pe}}$ and the inverse of the proper electron plasma frequency $\omega^{-1}_{\rm{pe}}$, respectively. Here 
\begin{equation}
    \omega_{\rm{pe}}=\sqrt{\frac{4\pi N_{0,\rm{e^-}}e^2}{\gamma_0 m_{\rm{e}}}},\label{eq:plasma_freq}
\end{equation}
where $N_{0,\rm{e^-}}$ is the upstream electron number density measured in the simulation frame, $\gamma_0$ is the bulk Lorentz factor of the upstream flow, and $m_e$ is the electron rest mass.

In the present study, we adopt the ion-to-electron mass ratio of $m_{\rm{i}}/m_{\rm{e}}=100$, the number of particles per cell in the upstream region of $100$, and the bulk Lorentz factor of $\gamma_0=40$. The number densities of positrons and ions are determined so that they satisfy the charge neutrality $N_{0,\rm{e^-}}=N_{0,\rm{e^+}}+N_{0,\rm{i}}$, where the subscripts denote particle species of electron ($\rm{e^{-}}$), positron ($\rm{e^{+}}$), and ion ($\rm i$), respectively. The computational cell size is set to $c/\omega_{\rm{pe}}/\Delta x=40$, and the number of cells in the $x$-direction is $160,000$. The time step size is $\omega_{\rm{pe}}\Delta t=1/40$. 

We introduce the magnetization parameter $\sigma_{\rm{s}}$, which is the ratio of the Poynting flux to the kinetic energy flux for each particle species as
\begin{equation}
    \sigma_{\rm{s}}\equiv\frac{B_0^2}{4\pi\gamma_0 N_{0,\rm{s}}m_{\rm{s}}c^2},\label{eq:sigma_particle}
\end{equation}
where $B_0$ is the magnetic field strength in the upstream region in the simulation frame. We also introduce the total magnetization, which is the sum of the inverse of each magnetization parameter, defined as
\begin{equation}
    \sigma_{\rm{tot}}\equiv\left(\frac{1}{\sigma_{\rm{e^-}}}+\frac{1}{\sigma_{\rm{e^+}}}+\frac{1}{\sigma_{\rm{i}}}\right)^{-1}.\label{eq:sigma_tot1}
\end{equation}
From Equation (\ref{eq:sigma_particle}), the total magnetization can be expressed with $\sigma_{\rm{e^-}}$ and 
$N_{0,\rm{e^+}}/N_{0,\rm{e^-}}$ as
\begin{equation}
    \sigma_{\rm{tot}}=\frac{\sigma_{\rm{e^-}}}{1+\frac{N_{0,\rm{e^+}}}{N_{0,\rm{e^-}}}+\frac{m_{\rm{i}}}{m_{\rm{e}}}\left(1-\frac{N_{0,\rm{e^+}}}{N_{0,\rm{e^-}}}\right)}.\label{eq:sigma_tot2}
\end{equation}
We adopt $\sigma_{\rm{tot}}=0.15$, which ensures maximum emission efficiency of the precursor waves via the synchrotron maser instability \citep{iwamotoPrecursorWaveAmplification2019}, and various positron fractions of 
$N_{0,\rm{e^+}}/N_{0,\rm{e^-}}=0,\ 0.05,\ 0.1,\ 0.2,\ 0.3,\ 0.35,\ 0.4$, and $0.5$ in the following PIC simulation runs.

\section{PIC simulation results} \label{sec:PIC}

\begin{figure*}[ht!] 
\plotone{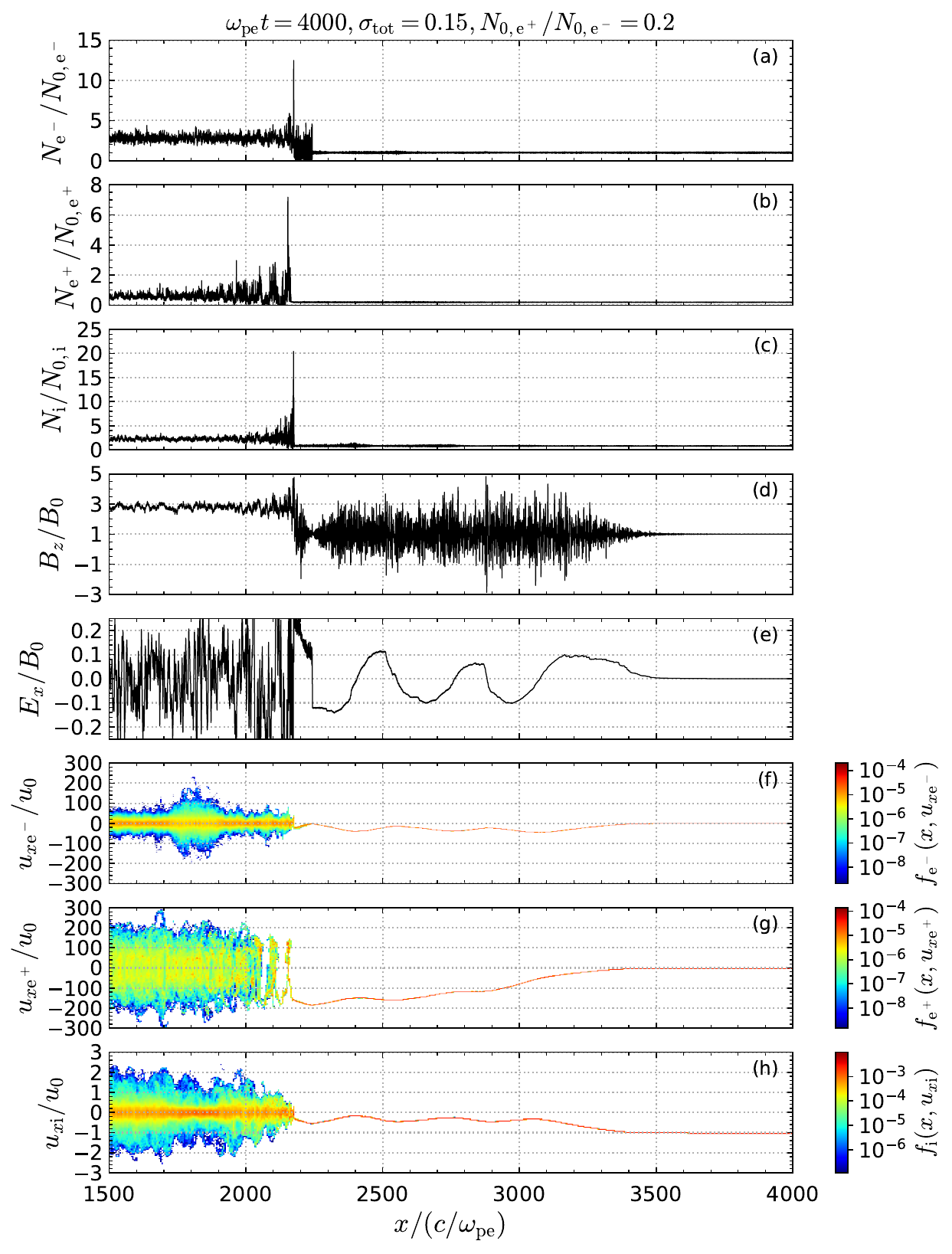}
\centering
\caption{Overall shock structure at $t=4000\ \omega_{\rm{pe}}^{-1}$ for $\sigma_{\rm{tot}}=0.15$ and $N_{0,\rm{e^+}}/N_{0,\rm{e^-}}=0.2$. From top to bottom, the number density profiles of (a) electron, (b) positron, and (c) ion, (d) the $z$-component of the magnetic field $B_z$, (e) the longitudinal electric field $E_x$, and the (f) electron, (g) positron, and (h) ion phase space densities $x-u_{x\mathrm{s}}$ are shown. All quantities are normalized to the upstream values, and the color scale of the phase-space plots is on a logarithmic scale.
\label{fig:ModerateSS}}
\end{figure*}
\begin{figure*}[ht!]
\plotone{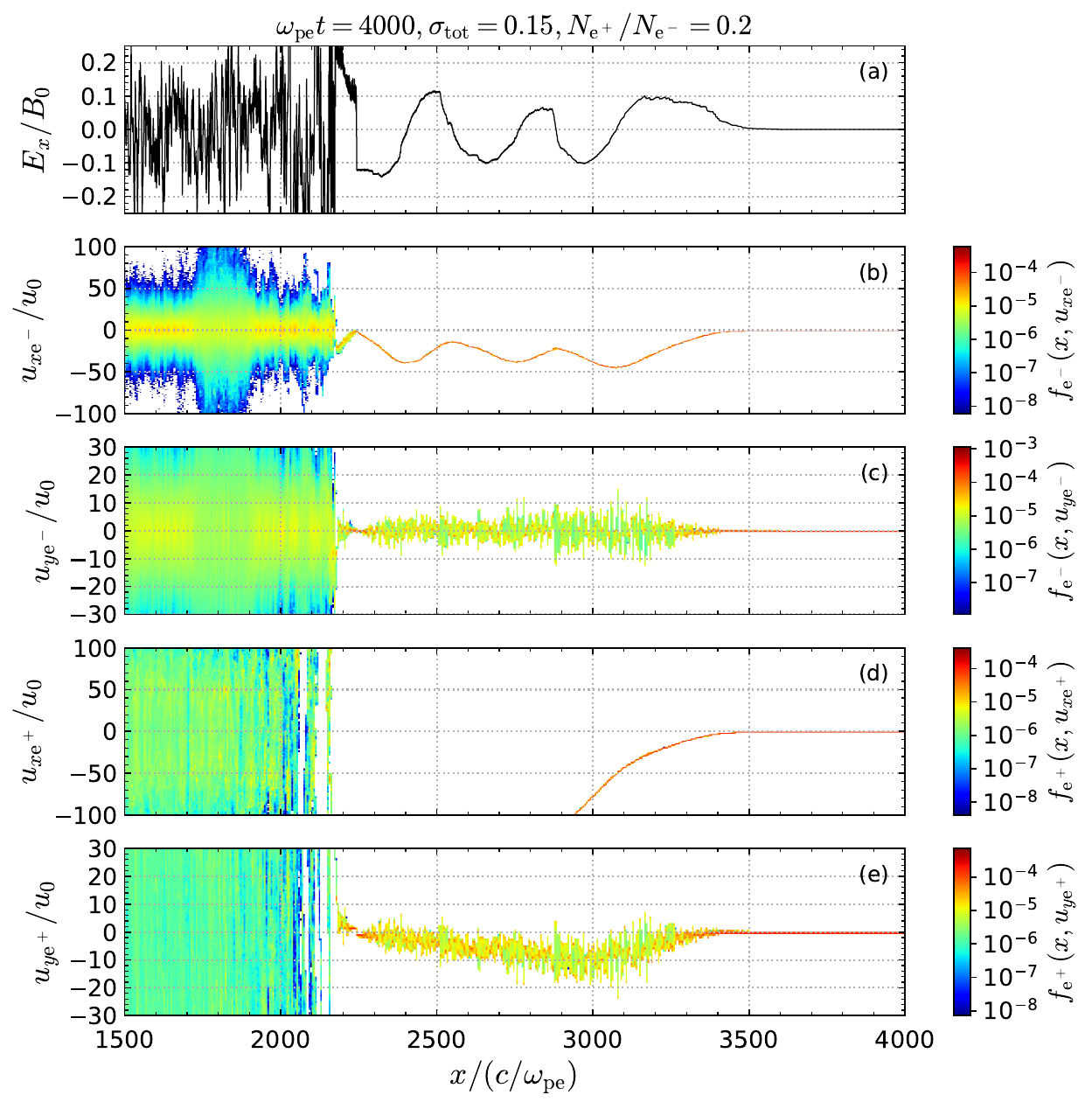}
\centering
\caption{Enlarged view of Figure \ref{fig:ModerateSS} around the wakefield in the upstream region. From top to bottom, shown are (a) $E_x$, electrons phase space densities for (b) $u_{xe^-}$ and (c) $u_{ye^-}$, and positron phase space densities for (d) $u_{xe^+}$ and (e) $u_{ye^+}$ in the same format as Figure \ref{fig:ModerateSS}.
\label{fig:PhaseSpaceSS}}
\end{figure*}
\begin{figure*}[ht!]
\plotone{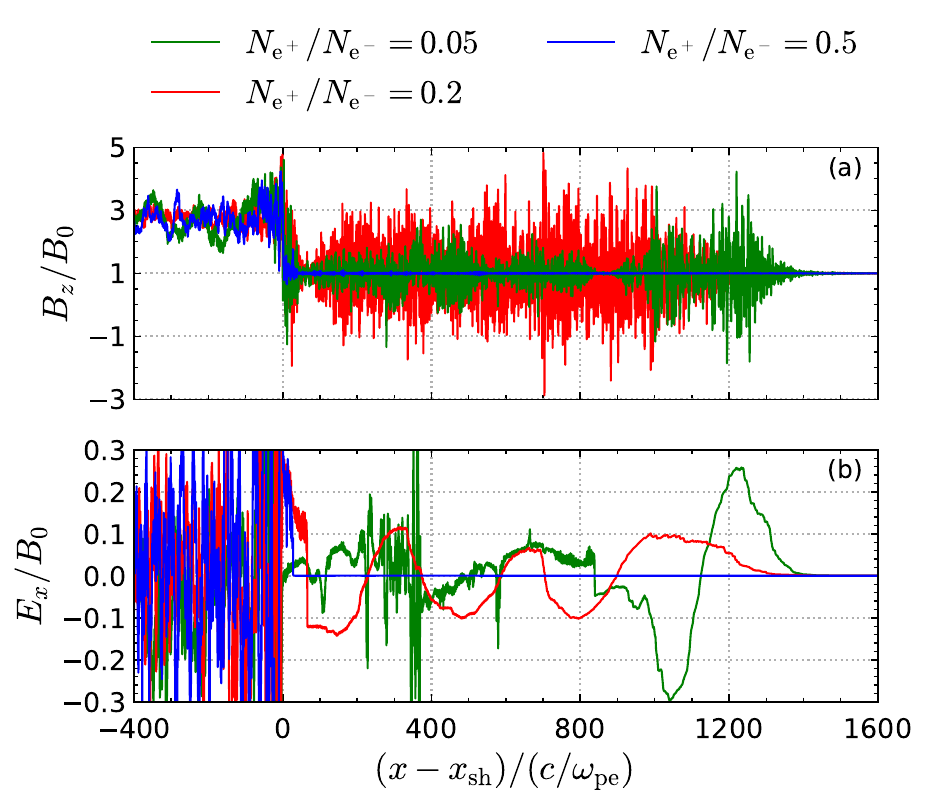}
\centering
\caption{(a) The $z$-component of the magnetic fields and (b) the $x$-component of the electric field for positron fractions of $N_{0,\rm{e^+}}/N_{0,\rm{e^-}}=0.05$ (green), $N_{0,\rm{e^+}}/N_{0,\rm{e^-}}=0.20$ (red), and $N_{0,\rm{e^+}}/N_{0,\rm{e^-}}=0.50$ (blue). $x-x_{\mathrm{sh}}$ indicates the position from the shock front. 
\label{fig:wave_wakefield}}
\end{figure*}
\begin{figure*}[ht!]
\plotone{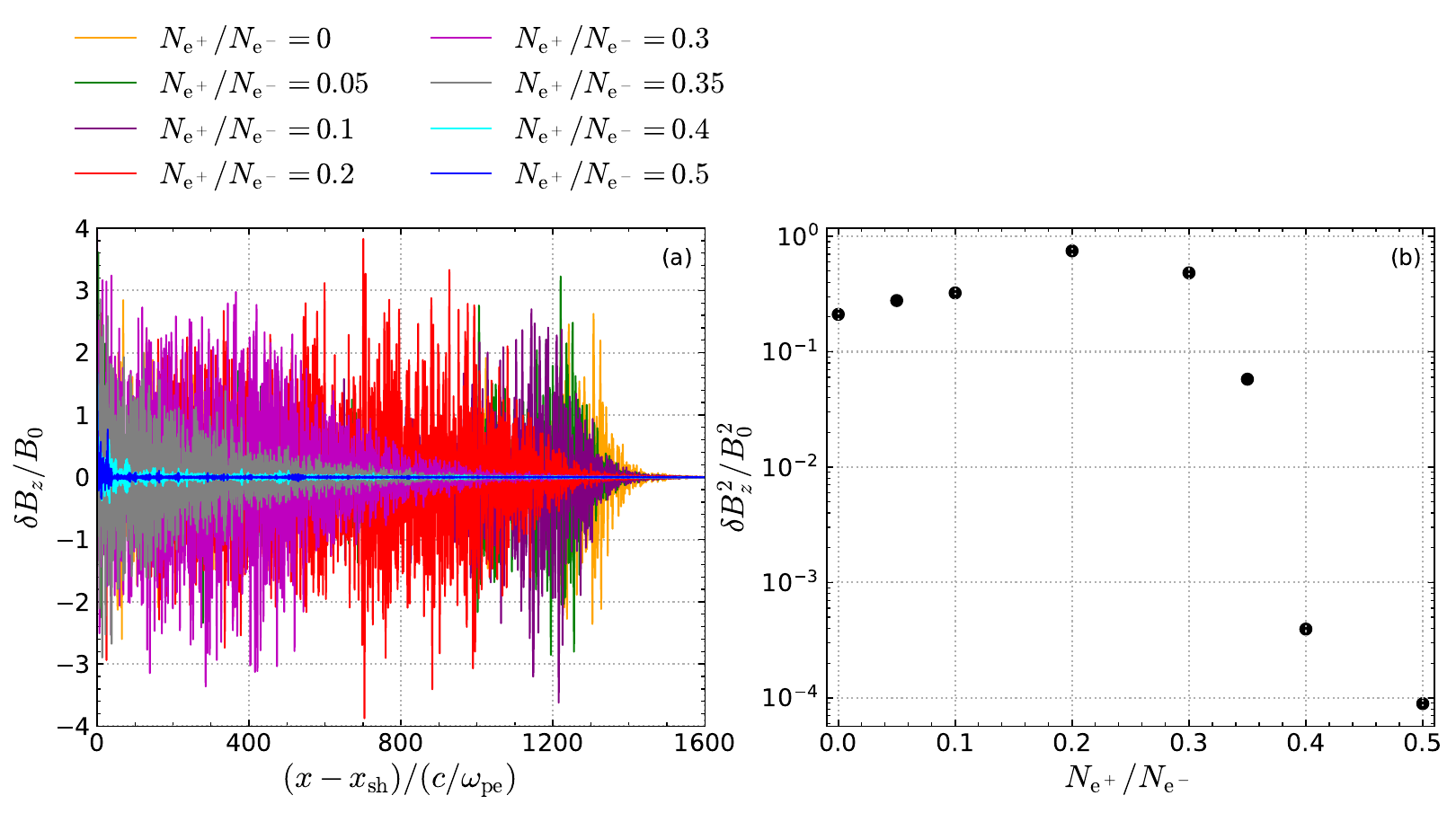}
\centering
\caption{(a) The electromagnetic wave emission and (b) the precursor wave energy averaged in $100 \leq (x-x_{\mathrm{sh}})/(c/\omega_{\mathrm{pe}}) \leq 1500$ for the various positron fractions.
\label{fig:elemag_wave}}
\end{figure*}
\begin{figure*}[ht!]
\plotone{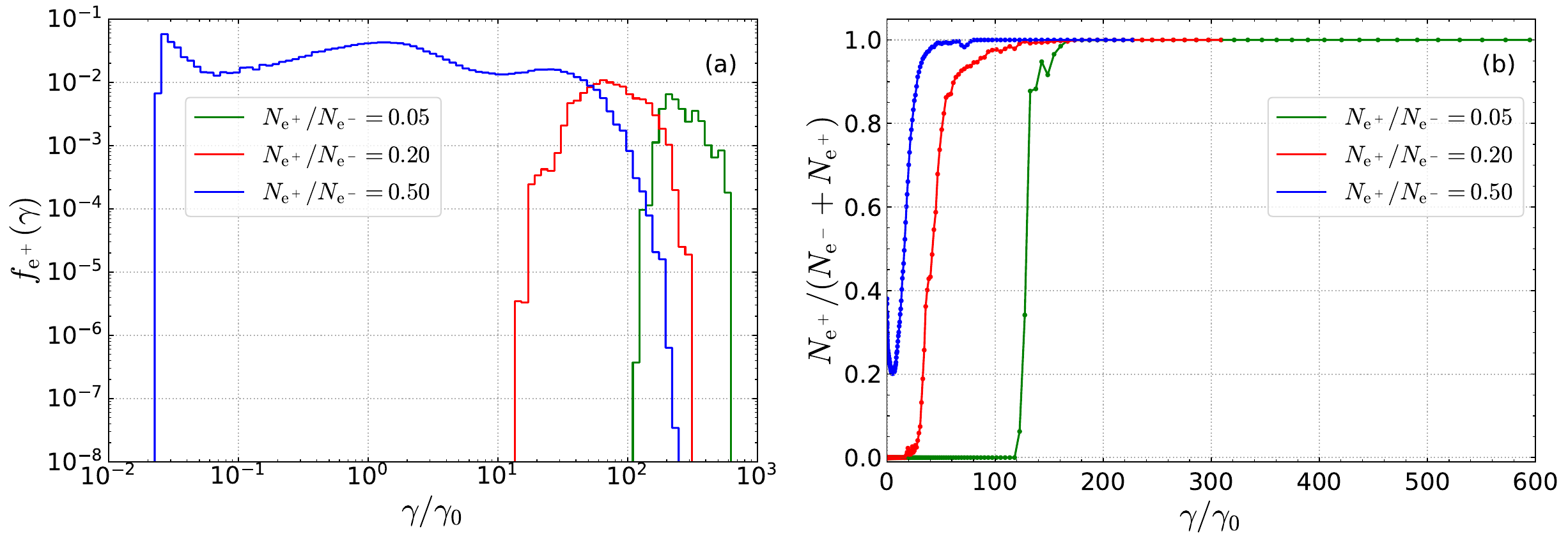}
\centering
\caption{(a) The downstream energy spectra of positrons for positron fractions of $N_{0,\rm{e^+}}/N_{0,\rm{e^-}}=0.05$ (green), $N_{0,\rm{e^+}}/N_{0,\rm{e^-}}=0.20$ (red), and $N_{0,\rm{e^+}}/N_{0,\rm{e^-}}=0.50$ (blue). (b) The ratio of the number of positrons to the sum of the number of electrons and positrons as a function of $\gamma$ for the various positron fractions.
\label{fig:spectra}}
\end{figure*}
Figure \ref{fig:ModerateSS} illustrates the global shock structure at $t=4000\ \omega_{\rm{pe}}^{-1}$ for $N_{0,\rm{e^+}}/N_{0,\rm{e^-}}=0.2$ case. At this time, the shock front is located at 
$x \sim 2180\ c/\omega_{\mathrm{pe}}$.
One can identify large-amplitude electromagnetic waves propagating upstream, followed by long-wavelength longitudinal electric fields. The latter is the so-called wakefield as already discussed in ion-electron shocks \citep[e.g.,][]{hoshinoWakefieldAccelerationRadiation2008}. 

From Figure \ref{fig:ModerateSS}(g), we found that positrons are preferentially accelerated during interactions with the wakefield. Figure \ref{fig:PhaseSpaceSS} shows an enlarged view around the wakefield to highlight the differences in particle accelerations between electrons and positrons. Electrons are accelerated and decelerated in response to changes in the sign of $E_x$ (Figures \ref{fig:PhaseSpaceSS}(b) and \ref{fig:PhaseSpaceSS}(c)). On the other hand, during the early stage of the interaction, positrons do not respond to the sign of $E_x$. They are accelerated even when $E_x$ is positive, as shown in Figure \ref{fig:PhaseSpaceSS}(d). Additionally, positrons are accelerated in the $y$-direction in
$3000 \lesssim x/(c/\omega_{\mathrm{pe}}) \lesssim 3400$
and are decelerated in
$2200 \lesssim x/(c/\omega_{\mathrm{pe}}) \lesssim 3000$.
In contrast, electrons exhibit fluctuations across $u_y = 0$, reflecting quiver motions induced by the precursor waves. Therefore, we cannot attribute the preferential acceleration of positrons solely to simple wakefield acceleration. It appears that a new mechanism for the acceleration is necessary. We will explore this further in the next section.

Figure \ref{fig:wave_wakefield}(a) and \ref{fig:wave_wakefield}(b) show the $z$-component of the magnetic field and the wakefield component for the various positron fractions, respectively. When $N_{0,\rm{e^+}}/N_{0,\rm{e^-}}=0.05$, the wakefield breaks up and becomes stochastic, similar to those found in electron-ion shocks \citep{hoshinoWakefieldAccelerationRadiation2008,iwamotoPrecursorWaveAmplification2019}. When $N_{0,\rm{e^+}}/N_{0,\rm{e^-}}=0.2$, on the other hand, a steady amplitude of electromagnetic waves is emitted over an extended period, and a sinusoidal wakefield continues to be formed. Neither the strong electromagnetic wave nor the wakefield is generated in the $N_{0,\rm{e^+}}/N_{0,\rm{e^-}}=0.5$ case.

Figure \ref{fig:elemag_wave}(a) focuses on the precursor waves in various positron fractions. Considering that electromagnetic waves observed further from the shock front were emitted at an earlier time, Figure \ref{fig:elemag_wave}(a) illustrates that a lower positron fraction is correlated with a more rapid growth rate for the emitted electromagnetic waves. Figure \ref{fig:elemag_wave}(b) shows the emitted wave energies for different positron fractions. The precursor wave energies are averaged in $100 \leq (x-x_{\mathrm{sh}})/(c/\omega_{\mathrm{pe}}) \leq 1600$. As a result, the wave energy peaks at $N_{\mathrm{e^+}}/N_{\mathrm{e^-}}=0.2$ and decreases for the fractions in $0.3\leq N_{\mathrm{e^+}}/N_{\mathrm{e^-}}$. We found that the accelerated positron significantly contributed to the overall emission efficiency around $N_{\mathrm{e^+}}/N_{\mathrm{e^-}}=0.2$.

Figure \ref{fig:spectra}(a) shows the energy spectra for positrons in $200~c/\omega_{\mathrm{pe}}$-wide areas located at $500~c/\omega_{\mathrm{pe}}$ downstream from the shock front at \(t = 4000\ \omega_{\rm{pe}}^{-1}\) for the various positron fractions. The spectrum is represented as a probability density function $f_{\rm{e^{+}}}(\gamma)$ defined as $\int_1^\infty f_{\rm{e^{+}}}(\gamma) d\gamma = 1$. With $N_{0,\rm{e^+}}/N_{0,\rm{e^-}}=0.5$, the thermal peak is found at the upstream positron kinetic energy where $\gamma\sim \gamma_0$, and a nonthermal population extends up to $\gamma\sim 100\gamma_0$, indicating an ion-positron coupling. The overall energy spectrum resembles that reported by \cite{hoshinoRelativisticMagnetosonicShock1992}. In the cases with $N_{0,\rm{e^+}}/N_{0,\rm{e^-}}=0.2$ and $N_{0,\rm{e^+}}/N_{0,\rm{e^-}}=0.05$, the energy spectra show that most positrons are accelerated to upstream ion kinetic energy, $\gamma \sim 100\gamma_0=\frac{m_{\rm{i}}}{m_{\rm{e}}}\gamma_0$, or even higher energies in the $N_{0,\rm{e^+}}/N_{0,\rm{e^-}}=0.05$ case. Additionally, as the upstream positron fraction decreases, the peak in the spectrum shifts toward higher energies. These results suggest a stronger coupling between ions and positrons in small positron fraction cases. The narrow width of the downstream distribution suggests that the positrons undergo a ballistic acceleration in the upstream region due to the wakefield, which is also supported by the phase space density shown in Figure \ref{fig:ModerateSS}(g). The particles are not fully isotropized in the downstream region in the present 1D simulations.

Figure \ref{fig:spectra}(b) shows the population fraction of positrons relative to the total population of electrons and positrons across various energy ranges. In all cases, we found that positrons dominated the population when the energy exceeded the upstream ion kinetic energy (\(\gamma > 100\gamma_0\)). As shown in Figure \ref{fig:spectra}(a), a smaller upstream positron fraction results in a higher energy positron dominance even though they constitute a minor population in the upstream region.

\section{Relativistic $\boldsymbol{E}\times\boldsymbol{B}$ Motion by Test Particle Simulations} \label{sec:TestParticle}
\begin{figure*}[ht!]
\plotone{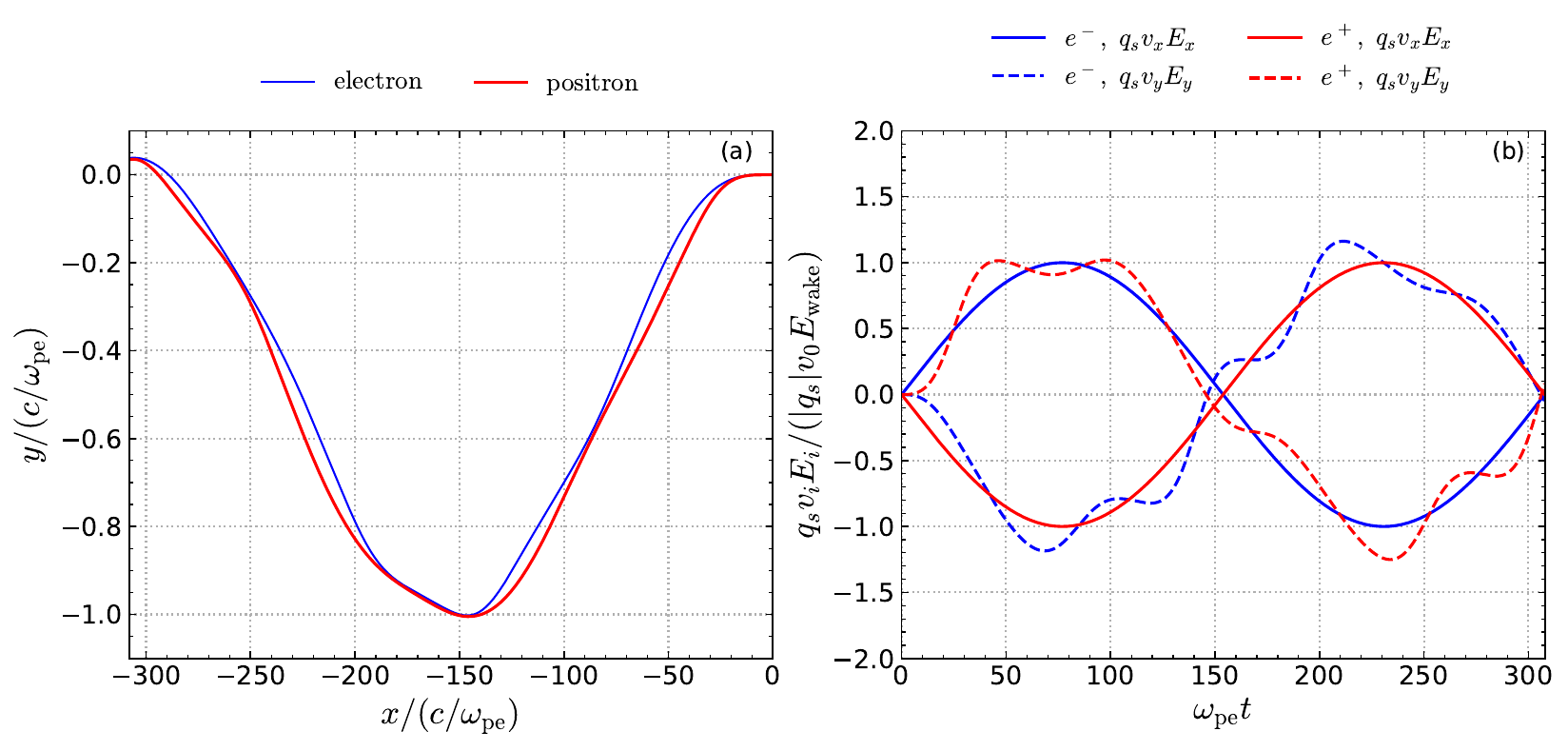}
\centering
\caption{The results from  test particle simulations with $E_{\rm{wake}}/B_0=0.01$. (a) The electron (solid blue line) and positron (solid red line) trajectories in the $x$--$y$ plane. (b) The time variation of the work rate done on particles. The $x$- and $y$-components of $q_sv_{s,i}E_i$, where $q_s$ and $v_{s,i}$ are respectively the charge and the $i$-component of velocity of particle species $s$, are shown with solid and dashed lines, respectively.
\label{fig:test_particle_plot1}}
\end{figure*}
\begin{figure*}[ht!]
\plotone{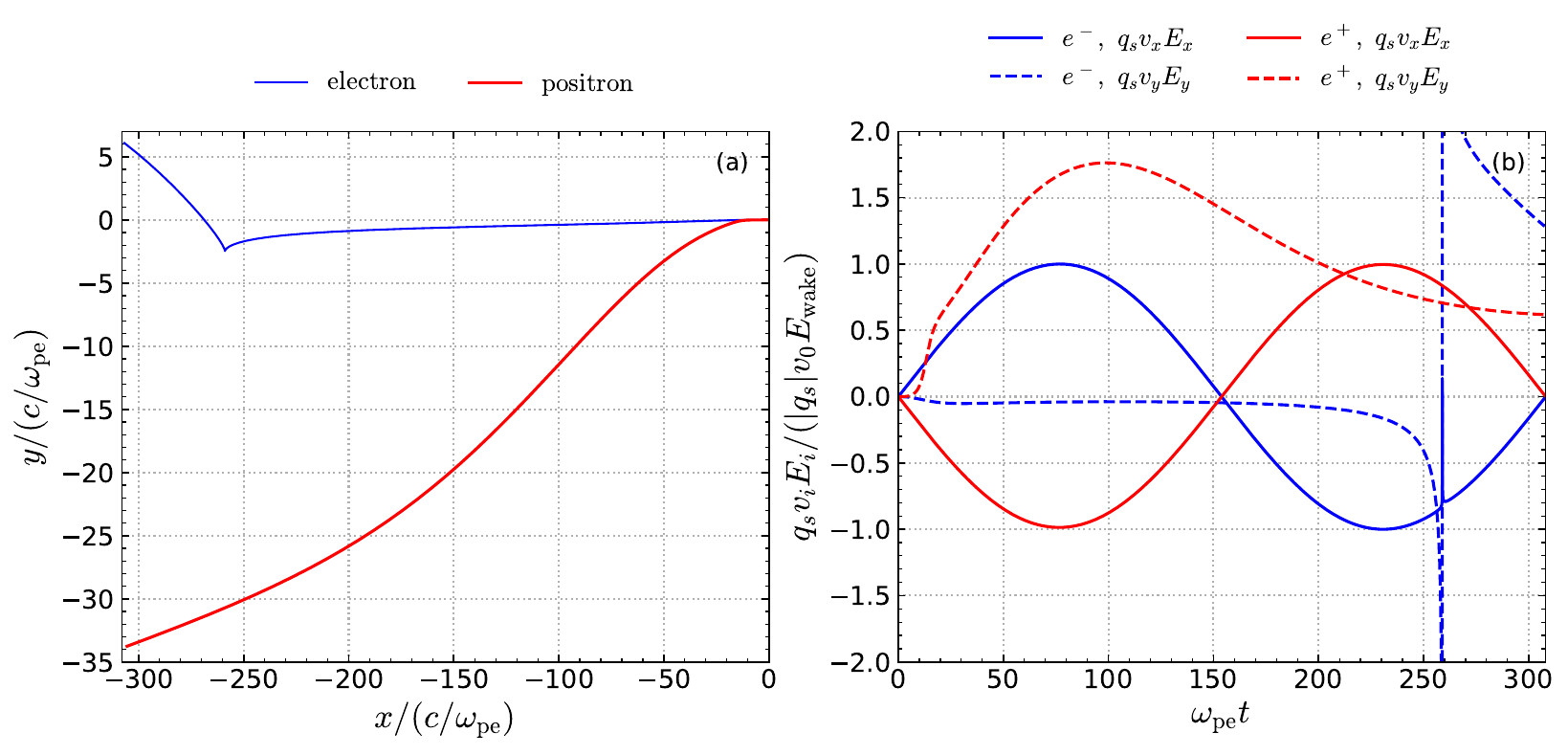}
\centering
\caption{The results from  test particle simulations with $E_{\rm{wake}}/B_0=0.1$ in the same format as Figure \ref{fig:test_particle_plot1}.
\label{fig:test_particle_plot2}}
\end{figure*}
We conducted test particle simulations to understand the asymmetric motions between electrons and positrons. We considered the ambient magnetic field $B_0$ pointing in the positive $z$-direction, the motional electric field $E_y=-\beta_0B_0$ with $\beta_0 = \sqrt{1-1/\gamma_0^2}$, and the wakefield in the $x$-direction. The wakefield propagating in the positive $x$-direction is modeled as
\begin{equation}
    E_x=\left\{
    \begin{array}{ll}
    -E_{\rm{wake}}\sin{\left[\frac{2\pi}{\lambda_{\rm{wake}}}(x-v_{\rm{wake}}t)\right]} & (v_{\rm{wake}}t-\lambda_{\rm{wake}} < x <v_{\rm{wake}}t)\\
    0 & (x\leq v_{\rm{wake}}t-\lambda_{\rm{wake}},\ v_{\rm{wake}}t \leq x )
    \end{array}.
    \right. \label{eq:Ex}
\end{equation}
where $E_{\rm{wake}}$ is the amplitude, $\lambda_{\rm{wake}}$ is the wavelength, and $v_{\rm{wake}}$ is the phase speed of the wakefield in the downstream rest frame. Based on the PIC simulation results, we selected $\lambda_{\rm{wake}}=600\ c/\omega_{\rm{pe}}$, and $v_{\rm{wake}}=+0.95~c$. Particles are injected at $x=0$ in the $-x$ direction with an initial Lorentz factor of $\gamma_0=40$. 
The adopted wavelength of the wakefield is determined by the one in $2880 \lesssim x/(c/\omega_{\rm{pe}}) \lesssim 3480$ in Figure 1(e) to model the upstream plasma interaction with the first sinusoidal wave.
Their motions are tracked in the $x$--$y$ plane according to the relativistic equations of motion until $t \sim 300\ \omega_{\rm{pe}}^{-1}$, at which the particles have experienced the entire phase of the sinusoidal wave. We examined two simulation runs; one with $E_{\rm{wake}}/B_0=0.01$ ($E_{\rm{wake}}/B^\prime_0=0.4$ in the plasma rest frame) and another with $E_{\rm{wake}}/B_0=0.1$ ($E_{\rm{wake}}/B^\prime_0=4.0$ in the plasma rest frame) to model the PIC simulation results.

Figure \ref{fig:test_particle_plot1} shows results from the run with $E_{\rm{wake}}/B_0=0.01$. As shown in Figure \ref{fig:test_particle_plot1} (a), an electron's (blue) and a positron's (red) trajectories coincide with each other with a slight excursion in the $-y$-direction. Figure \ref{fig:test_particle_plot1}(b) illustrates the time variation of the work rate done on the particles by the electric fields. Although the work rate fluctuates following the sign of the wakefield, the overall energy gain after completing the entire phase is nearly zero. This indicates that both the electron and the positron experience standard $\boldsymbol{E}\times\boldsymbol{B}$ drift motion in the presence of the wakefield. 

When $E_{\rm{wake}}/B_0=0.1$, it results in asymmetric motions between the electron and the positron as shown in Figure \ref{fig:test_particle_plot2}. The motions of the electron and the positron no longer coincide, causing their positions in the $x$--$y$ plane to diverge over time. The electron primarily drifts in the $-x$-direction and then experiences a sudden change in the $y$ direction at $x\approx-260\ c/\omega_{\rm{pe}}$. This change can be understood by a pick-up process by the motional electric field when the particle's speed approaches zero \citep{iwamotoParticleAccelerationPickup2022}. On the other hand, the positron drifts not only in the $x$-direction but also in the $-y$-direction, consistent with the PIC simulation results (Figure \ref{fig:PhaseSpaceSS}).

The difference in the motion of the particles can be understood by examining the work rate shown in Figure \ref{fig:test_particle_plot2}(b). The work rate in the $x$-direction reflects the variation of the wakefield, and the net work on both particles is almost zero. This implies that the wakefield does not directly influence their acceleration in this reference frame. Interestingly, the motional electric field accelerates the positron, as evidenced by the positive net work in the $y$-direction (red dashed line). In contrast, the electron gains little energy in this direction until it enters the pick-up process. After the pick-up process, the net work becomes positive, indicating the acceleration (blue dashed line). Note, however, that we have not observed this pick-up process in the present 1D PIC simulations.

The results of test particle simulations suggest that the amplitude of the wakefield plays a crucial role in the preferential positron acceleration; under sufficiently large amplitudes of the wakefield, particles do not follow the standard $\boldsymbol{E}\times\boldsymbol{B}$ drift motion. We attribute this behavior to the relativistic $\boldsymbol{E}\times\boldsymbol{B}$ acceleration. \cite{takeuchiRelativistic$EifmmodetimeselsetexttimesfiB$Acceleration2002} and \cite{friedmanRelativisticAccelerationCharged2005} provided analytic solutions for the motion of a charged particle in uniform magnetic and electric fields that are orthogonal to one another. While the solution gives the standard drift motion in the non-relativistic electric field, $|E| < |B|$, the solution with $|E| \geq |B|$ predicts that the particle experiences continuous acceleration in different directions depending on the sign of its charge. 

The theory can be applied to the current situation by Lorentz transforming from the downstream rest (simulation) frame to the upstream rest frame. Since the $x$-component of the electric field is invariant in the Lorentz transformation to the upstream rest frame, $E_{\rm{wake}}/(B_0/\gamma_0)$ determines whether the relativistic $\boldsymbol{E}\times\boldsymbol{B}$ acceleration works effectively in the upstream rest frame. In fact, the ratio measured in the upstream rest frame adopted in the test particle simulations characterized different behaviors of the particles. In this frame, the positron (electron) can be accelerated in the positive (negative) $x$-direction by interacting with the sinusoidal wakefield propagating in the positive direction, resulting in the asymmetric motions of positrons and electrons in the downstream rest frame. In the next section, we focus on the quantity $E_{\rm{wake}}/(B_0/\gamma_0)$ as a measure of the preferential positron acceleration in various astrophysical situations.

\section{Preferential Positron Acceleration in Astrophysical Objects} \label{sec:accel_conditions}
\subsection{Condition for the $\boldsymbol{E}\times\boldsymbol{B}$ Acceleration}\label{subsec:ExBcondition}
To evaluate the efficiency of the relativistic $\boldsymbol{E}\times\boldsymbol{B}$ acceleration in relativistic magnetized shocks, we first estimate the wakefield amplitude in the downstream rest frame as \citep[cf.][]{hoshinoWakefieldAccelerationRadiation2008}
\begin{equation}
    \frac{E_{\rm{wake}}}{B_0}\simeq\frac{\eta a_0^2}{\sqrt{1+\eta a_0^2}}\left( \frac{1}{\sqrt{\sigma_{\rm{e^-}}}\gamma_0} \right),  \label{eq:E_wake1}
\end{equation}
where $a_0=\frac{eE_{\rm{EM}}}{m_{\rm{e}}c\omega_{\rm{EM}}}$ is the strength parameter, and $E_{\rm{EM}}$ and $\omega_{\rm EM}$ refer to the electric field amplitude and the angular frequency of the precursor electromagnetic wave, respectively. Since the precursor wave has a strong linear polarization property \citep{iwamotoLinearlyPolarizedCoherent2024}, we may set $\eta=\frac{1}{2}$.

We should take into account the positron response to the ponderomotive force by multiplying Equation (\ref{eq:E_wake1}) by the fraction of the ion to the electron as
\begin{equation}
    \frac{E_{\rm{wake}}}{B_0}\approx\frac{N_{0,\rm{i}}}{N_{0,\rm{e^-}}}\frac{\eta a_0^2}{\sqrt{1+\eta a_0^2}}\left( \frac{1}{\sqrt{\sigma_{\rm{e^-}}}\gamma_0} \right).  \label{eq:E_wake2}
\end{equation}
This modification guarantees that the wakefield disappears in the limit of the pair plasma case.

\begin{figure*}[ht!]
\plotone{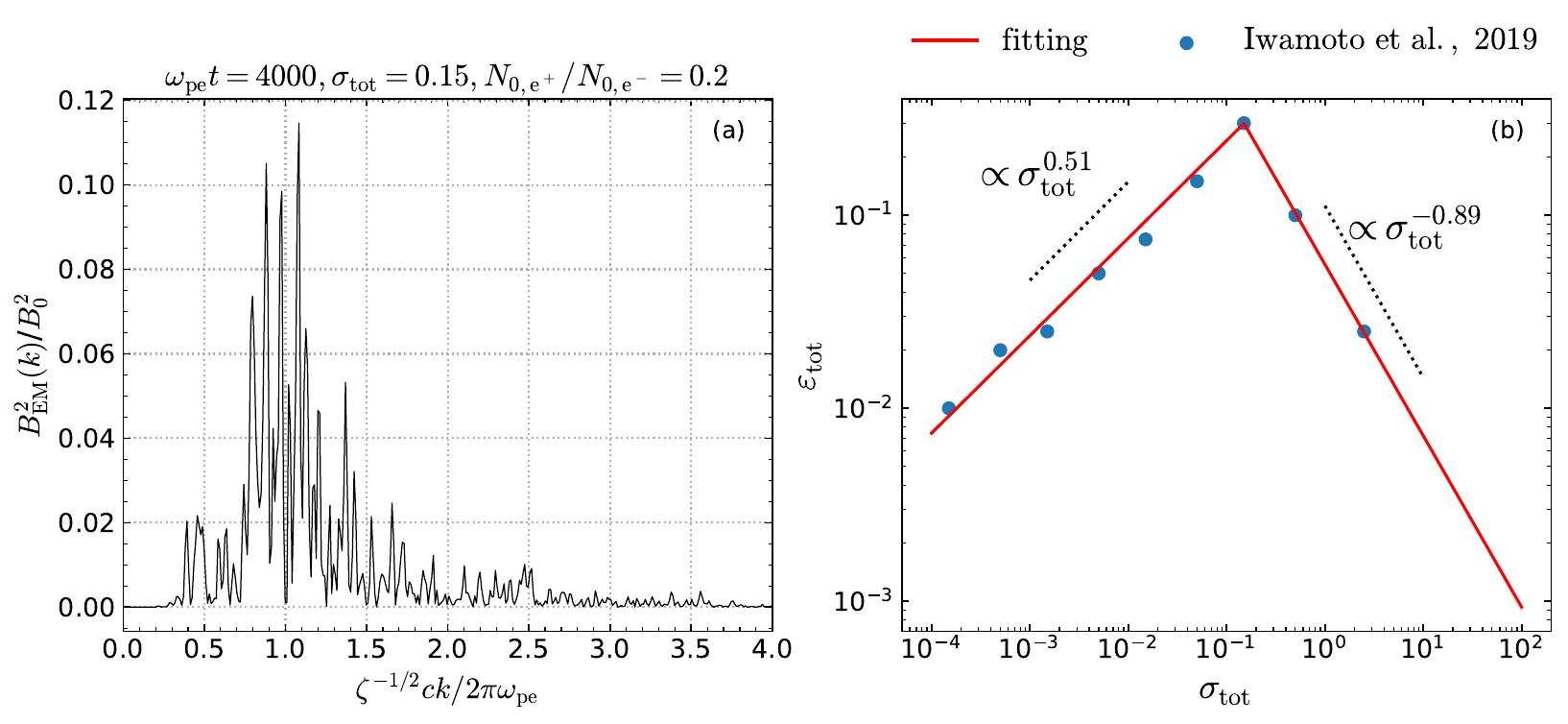}
\centering
\caption{(a) The wave power spectrum of the wave component of the magnetic field $\delta B_z=B_z-B_0$ normalized by upstream ambient magnetic field in the upstream region in
$2880 < x/(c/\omega_{\rm{pe}}) <3480$
at $t=4000~\omega_{\rm{pe}}^{-1}$.The abscissa is the wavenumber normalized by $c/\omega_{\rm{pe}}$ multiplied by $\zeta^{-\frac{1}{2}}$ which is the factor of the electron-positron-ion coupling (Equation (\ref{eq:zeta})). (b) The precursor wave energy conversion rate $\epsilon_{\rm{tot}}$ to the total upstream kinetic energy as a function of $\sigma_{\rm{tot}}$. Data in the panel is adopted from \cite{iwamotoPrecursorWaveAmplification2019}. The data fit results for $\sigma_{\rm tot} \le 0.15$ and $\sigma_{\rm tot} < 0.15$ are shown with solid red lines. The dotted lines indicate $\epsilon_{\rm{tot}}\propto\sigma_{\rm{tot}}^{0.51}$ and $\epsilon_{\rm{tot}}\propto\sigma_{\rm{tot}}^{-0.89}$, respectively.}
\label{fig:fourier_fitting}
\end{figure*}

Next we estimate $\omega_{\rm{EM}}$ and $E_{\rm{EM}}$, and thus $a_0$, from the simulation results as follows. Figure \ref{fig:fourier_fitting}(a) shows the wave power spectrum of the wave component of the magnetic field ($\delta B_z = B_z-B_0$). This spectrum was obtained from a snapshot at $t=4000~\omega_{\rm{pe}}^{-1}$ by applying the Hanning window and taking the Fourier transformation in the upstream region of
$2880 < x/(c/\omega_{\rm{pe}}) <3480$.
In Figure \ref{fig:fourier_fitting}(a), we introduce a factor
\begin{equation}
    \zeta=\frac{2m_{\rm{e}}}{m_{\rm{i}}\left[ 2+\left( m_{\rm{e}}/m_{\rm{i}}-1 \right)\left( N_{0,\rm{e^+}}/N_{0,\rm{e^-}}+1 \right) \right]}, \label{eq:zeta}
\end{equation}
which represents the electron-positron-ion coupling. From Figure \ref{fig:fourier_fitting}(a), we found that the peak wave number in the downstream rest frame can be expressed
\begin{equation}
    k_{\rm{EM}}\sim2\pi\times\zeta^{1/2}\frac{\omega_{\rm{pe}}}{c} \approx \zeta^{1/2}\frac{\omega_{\rm{pe}}}{c} , \label{eq:k_EM}
\end{equation}
consistent with the case for the electron-ion coupling \citep{lyubarskyEmissionMechanismsFast2021}. The dispersion relation of the extraordinary mode in the downstream rest frame can be written as \citep{ligoriniMildlyRelativisticMagnetized2021}
\begin{equation}
    \omega_{\rm{EM}} = \sqrt{\left(1+\sigma_{\rm{e^-}}/\gamma^2_0\right)\omega^2_{\rm{pe}}+c^2k^2},
\end{equation}
which gives the angular frequency corresponding to the wavenumber (Equation (\ref{eq:k_EM})) as
\begin{equation}
    \omega_{\rm{EM}}\sim\sqrt{1+\frac{\sigma_{\rm{e^-}}}{\gamma^2_0}+\zeta}\ \omega_{\rm{pe}}. \label{eq:omega_EM}
\end{equation}

Next, we estimate $E_{\rm{EM}}$ assuming $E_{\rm{EM}} \approx B_{\rm{EM}}$ from the precursor wave energy conversion rate to the total upstream kinetic energy $\epsilon_{\rm{tot}}$, 
\begin{equation}
    \epsilon_{\rm{tot}}\equiv\frac{B^2_{\rm{EM}}}{4\pi\left[ (N_{0,\rm{e^-}}+N_{0,\rm{e^-}})m_{\rm{e}}+N_{0,\rm{i}}m_{\rm{i}} \right]c^2}=\sigma_{\rm{tot}}\frac{B^2_{\rm{EM}}}{B_0^2}.  \label{eq:epsilon_tot1}
\end{equation}
Figure \ref{fig:fourier_fitting}(b) shows the conversion rate as a function of $\sigma_{\rm{tot}}$ from 1D PIC simulations (blue dots) \citep{iwamotoPrecursorWaveAmplification2019}, and their best fit results (solid red lines). The fitting curves are expressed as
\begin{equation}
    \epsilon_{\rm{tot}}\approx \left\{
    \begin{array}{ll}
    0.783~\sigma_{\rm{tot}}^{0.505} & (\sigma_{\rm{tot}}\leq0.15),\\
    0.0557~\sigma_{\rm{tot}}^{-0.888} & (\sigma_{\rm{tot}}>0.15),
    \end{array}
    \right. \label{eq:epsilon_tot2}
\end{equation}
one of which in $\sigma_{\rm{tot}}>0.15$ is consistent with the previous results \citep{plotnikovSynchrotronMaserEmission2019}. Combining Equations (\ref{eq:epsilon_tot1}) and (\ref{eq:epsilon_tot2}), we obtain the amplitude of the electromagnetic waves as a function of $\sigma_{\rm{tot}}$ as,
\begin{equation}
    \frac{E_{\rm{EM}}}{B_0}\approx
    \frac{B_{\rm{EM}}}{B_0}\approx \left\{
    \begin{array}{ll}
    0.88~\sigma_{\rm{tot}}^{-0.25} & (\sigma_{\rm{tot}}\leq0.15),\\
    0.24~\sigma_{\rm{tot}}^{-0.94} & (\sigma_{\rm{tot}}>0.15).
    \end{array}
    \right. \label{eq:E_EM}
\end{equation}
Using Equations (\ref{eq:E_wake2}), (\ref{eq:omega_EM}), (\ref{eq:E_EM}), and $a_0=eE_{\rm{EM}}/m_{\rm{e}}c\omega_{\rm{EM}}$, we obtain a condition for the preferential positron acceleration $E_{\rm{wake}}/(B_0/\gamma_0)>1$ in terms of the bulk Lorentz factor $\gamma_0$ and the total magnetization parameter $\sigma_{\rm tot}$ as
\begin{equation}
    \frac{\gamma_0 E_{\rm wake}}{B_0} = 0.62~\gamma_0\frac{N_{0,\rm{i}}}{N_{0,\rm{e^-}}}\left(1+\frac{\sigma_{\rm{e^-}}}{\gamma^2_0}+\zeta\right)^{-\frac{1}{2}}
    \sigma_{\rm{tot}}^{-0.25}
    > 1 ~~(\sigma_{\rm{tot}}\leq0.15),
    \label{eq:condition_of_acceleration1}
\end{equation}
and
\begin{equation}
    \frac{\gamma_0 E_{\rm wake}}{B_0} = 0.17~\gamma_0\frac{N_{0,\rm{i}}}{N_{0,\rm{e^-}}}\left(1+\frac{\sigma_{\rm{e^-}}}{\gamma^2_0}+\zeta\right)^{-\frac{1}{2}}
    \sigma_{\rm{tot}}^{-0.94}
    > 1 ~~(\sigma_{\rm{tot}}>0.15),
    \label{eq:condition_of_acceleration2}
\end{equation}
where we have assumed $\eta a_0^2\gg1$ and $\eta=\frac{1}{2}$. The marginal condition of Equations (\ref{eq:condition_of_acceleration1}) and (\ref{eq:condition_of_acceleration2}) is indicated by the solid line in Figure \ref{fig:energy_gains} for different $\frac{N_{0,\rm{i}}}{N_{0,\rm{e^-}}}$ using the real mass ratio of $\frac{m_{\rm{i}}}{m_{\rm{e}}}=1836$.

\subsection{Energy Gain}
\begin{figure*}[ht!]
\plotone{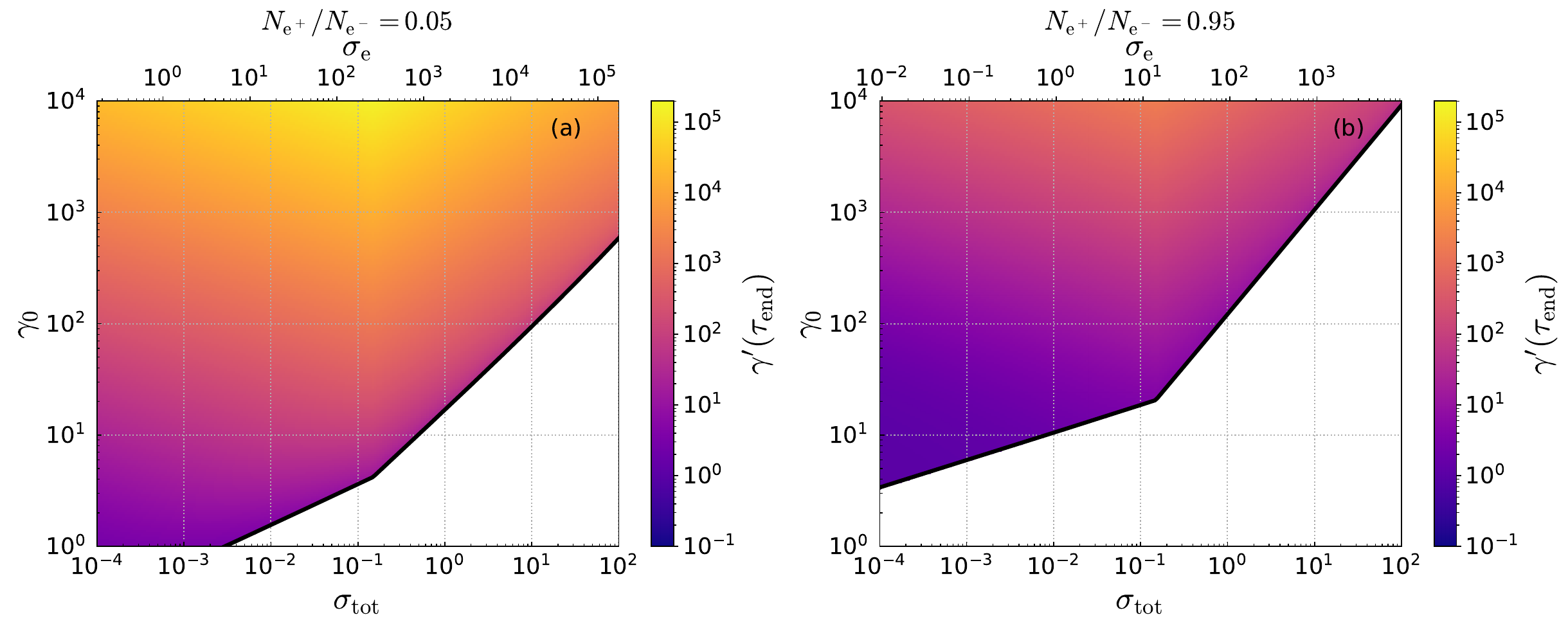}
\centering
\caption{Energy gain (Equation (\ref{eq:gamma_up2})) after the interaction with the wakefield in the upstream rest frame (color) as a function of the total (electron) magnetization (bottom horizontal (top horizontal) axis) and the upstream bulk Lorentz factor (vertical axis) for (a) $N_{\rm{e^+}}/N_{\rm{e^-}}=0.05$ and (b) $N_{\rm{e^+}}/N_{\rm{e^-}}=0.95$. The white area separated by the black solid line in each panel indicates the parameter space where the relativistic $\boldsymbol{E}\times\boldsymbol{B}$ acceleration is not expected according to Equations (\ref{eq:condition_of_acceleration1}) and (\ref{eq:condition_of_acceleration2}).
\label{fig:energy_gains}}
\end{figure*}

This subsection discusses the particle's energy gain after interacting with the wakefield. We consider constant, positive electric and magnetic field strengths of the $E_x$ and $B_z$ components in the upstream rest frame to discuss the energy gain order in magnitude. Hereafter, they are denoted by $E_0$ and $B_0^\prime$, respectively, and the prime symbol in superscript indicates the quantity in the upstream rest frame. We also assume a typical time scale for the acceleration as $t \sim \omega_{\rm pe}^{-1}$, which characterizes the Langmuir wave (the wakefield) oscillation period in the upstream rest frame. Given the uniform, constant fields and a particle initially at rest, \cite{friedmanRelativisticAccelerationCharged2005} provided an analytic form of the time evolution of the particle's Lorentz factor as
\begin{equation}
    \gamma^\prime=\gamma^{\prime 2}_{\rm{d}}\left[\cosh{(\nu^\prime\tau)}-\frac{v^{\prime 2}_{\rm{d}}}{c^2}\right],\label{eq:gamma_up}
\end{equation}
\begin{equation}
    t^\prime=\gamma^{\prime 2}_{\rm{d}}\left[\frac{\sinh{(\nu^\prime\tau)}}{\nu^\prime}-\frac{v^{\prime 2}_{\rm{d}}}{c^2}\tau\right],\label{eq:t_up}
\end{equation}
where $v^\prime_{\rm{d}}=cB^\prime_0/E_{0}$, $\gamma^\prime_{\rm{d}}=1/\sqrt{1-(v^\prime_{\rm{d}}/c)^2}$, and $\tau$ represents the proper time. Also, 
\begin{equation}
    \nu^\prime\equiv\frac{q_{\rm{s}}}{|q_{\rm{s}}|}\omega_{\rm{cs}}\sqrt{\left(\frac{E_{0}}{B_0^\prime}\right)^2-1}, \label{eq:nu_up}
\end{equation}
where $\omega_{\rm{cs}}=(|q_{\rm{s}}|B_0^\prime)/(m_{\rm{s}}c)$ is the cyclotron frequency of the particle. Equation (\ref{eq:gamma_up}) gives the final energy after a time of $\omega_{\rm pe}^{-1}$ as
\begin{equation}
    \gamma^\prime(\tau_{\rm end}) = \gamma^{\prime 2}_{\rm{d}}\left[\cosh{(\nu^\prime\tau_{\rm end})}-\frac{v^{\prime 2}_{\rm{d}}}{c^2}\right],\label{eq:gamma_up2}
\end{equation}
where $\tau_{\rm end}$ satisfies the relation
\begin{equation}
    \omega_{\rm pe}^{-1} = \gamma^{\prime 2}_{\rm{d}}\left[\frac{\sinh{(\nu^\prime\tau_{\rm end})}}{\nu^\prime}-\frac{v^{\prime 2}_{\rm{d}}}{c^2}\tau_{\rm end}\right].\label{eq:t_up2}
\end{equation}
We first solve Equation (\ref{eq:t_up2}) numerically to find $\tau_{\rm end}$, and then obtain $\gamma^\prime(\tau_{\rm end})$ analytically from Equation (\ref{eq:gamma_up2}) for different values of $\nu^\prime (E_0/B_0^\prime)\approx\nu^\prime (E_{\rm wake}/(B_0/\gamma_0))$. This exploration allows us to assess the energy gain under various conditions characterized by $\sigma_{\rm tot}$ and $\gamma_0$ from Equations (\ref{eq:condition_of_acceleration1}) and (\ref{eq:condition_of_acceleration2}).

Figure \ref{fig:energy_gains} shows the energy gain under various parameters of $\sigma_{\rm tot}$ and $\gamma_0$ for the two different positron fractions. In the ion-dominant case (Figure \ref{fig:energy_gains}(a)), the condition of $\boldsymbol{E} \times \boldsymbol{B}$ acceleration is satisfied across a broad range of parameters, except in the region characterized by sub-relativistic and highly magnetized conditions. For a given value of $\gamma_0$, the energy gain reaches its peak at $\sigma_{\rm tot}=0.15$, which corresponds to the maximum efficiency of the precursor wave emission (Figure \ref{fig:fourier_fitting}(b)). 

The maximum attainable energy from the instantaneous acceleration increases as the upstream bulk Lorentz factor increases, since the electric field becomes more prominent relative to the magnetic field strength measured in the upstream rest frame. The peak energy gain can exceed $\gamma = 10^5$ for $\gamma_0=10^4$.

In the positron-dominant case, as shown in Figure \ref{fig:energy_gains}(b), the necessary condition for the acceleration is still met in highly relativistic situations ($\gamma_0 > 100$). Although the energy gain is limited to $\gamma\sim100$ within the parameter space shown in Figure \ref{fig:energy_gains}(b), the mechanism could be essential in accelerating particles in ultra-relativistic shocks.

\section{Summary and Discussion} \label{sec:Discussion}
We have studied positron accelerations in relativistic magnetized electron-positron-ion shocks by examining 1D PIC simulations. We found that positrons are preferentially accelerated by interacting with the wakefield in the upstream region associated with precursor wave emission. The energy spectra in the downstream region reveal that, despite the relatively small upstream positron population, the high-energy positrons significantly outnumber the electrons. This preferential positron acceleration suggests that relativistic magnetized shocks could be a source of the high-energy primary positrons, potentially explaining the excess high-energy population as has been reported by the space experiments.

As shown in Figure \ref{fig:wave_wakefield}(b), a coherent, sinusoidal wakefield profile persists in $N_{\mathrm{e^+}}/N_{\mathrm{e^-}} = 0.2$ case. This can be understood by a contribution from the accelerated positron. Since positrons are efficiently and ballistically accelerated, they form a well-defined ring distribution as they enter the shock front, resulting in the emission of large-amplitude electromagnetic waves. In case of $N_{\mathrm{e^+}}/N_{\mathrm{e^-}} = 0.05$, however, the wakefield becomes rather stochastic around the shock front as has been shown in the electron-ion shocks. Positron's contribution to the overall emission efficiency is limited in such very small fraction cases. Since positrons accelerated in the far upstream region already had large enough energy, their motions are hardly affected by a turbulent wakefield even in the $N_{\mathrm{e^+}}/N_{\mathrm{e^-}} = 0.05$ case.

The gradual growth of the precursor wave amplitude shown in Figure \ref{fig:elemag_wave}(a) can be attributed to the positive feedback process \citep{lyubarskyElectronIonCouplingUpstream2006,hoshinoWakefieldAccelerationRadiation2008,iwamotoPrecursorWaveAmplification2019}. As the positron fraction exceeds $N_{\mathrm{e^+}}/N_{\mathrm{e^-}} = 0.2$, the amplitude of the wakefield decreases due to reduced charge separation, thereby weakening the positive feedback. This suggests a competing process between the positive contribution from the accelerated positron and screening of the wakefield by the surrounding positron. This competition results in the peak in the emission efficiency around $N_{\mathrm{e^+}}/N_{\mathrm{e^-}} = 0.2$. Nevertheless, the current simulations may not be sufficiently long, and further long-term simulations could demonstrate more efficient precursor emission for $N_{\mathrm{e^+}}/N_{\mathrm{e^-}} \geq 0.35$.

The efficiency of the precursor wave emission and resulting positron accelerations depend on the upstream plasma temperature. In electron-positron shocks with $\sigma_{\rm{e}} \gtrsim 1$, while the efficiency of the precursor wave is independent of the thermal spread as long as $k_BT/m_ec^2 \lesssim 10^{-1.5}$, where $k_B$ and $T$ are the Boltzmann constant and the upstream electron temperature, respectively, there is a significant decrease in efficiency in $k_BT/m_ec^2>10^{-1.5}$ \citep{babulSynchrotronMaserEmission2020}. Since the population inversion of the energy distribution is the free energy for the synchrotron maser instability, thermal spread of the distribution function is expected to suppress the instability growth. Although the physics of electron-positron-ion shocks differ qualitatively, such suppression could work in the electron-positron-ion case under hot upstream plasma conditions.

We attributed the positron acceleration to the relativistic $\boldsymbol{E} \times \boldsymbol{B}$ acceleration. Test-particle simulations revealed that the amplitude of the wakefield needs to be greater than the ambient magnetic field measured in the upstream rest frame to satisfy the condition of the relativistic $\boldsymbol{E} \times \boldsymbol{B}$ acceleration. We formulated the necessary condition for the acceleration in terms of the upstream bulk Lorentz factor and the total magnetization parameter. The results indicate that ultra-relativistic, mildly magnetized conditions, such as those found in pulsar winds, are likely effective sites for the positron preferential acceleration. In contrast, relativistic jets from active galactic nuclei, which are characterized by mildly relativistic shock speeds, are not anticipated to be efficient accelerators of high-energy primary positrons.

The present mechanism applies not only to astrophysical shocks but also to laser-plasma interactions in laboratory experiments, since the situations are fundamentally similar in the upstream rest frame. Laser experiments have reported that positrons can be accelerated by the wakefield self-generated by a positron beam in unmagnetized plasma \citep[e.g.,][]{bluePlasmaWakefieldAccelerationIntense2003, cordeMultigigaelectronvoltAccelerationPositrons2015}. If the wakefield is produced by a laser pulse in magnetized, electron-positron-ion plasma at rest, whose strength parameter satisfies the condition derived from Equation (\ref{eq:E_wake2}) as
\begin{equation}
a_0 > \sqrt{\sigma_{\rm{e^-}}},
\end{equation}
positrons would be accelerated similarly to what is observed in present shock simulations. The broader application of this mechanism will be explored and discussed in future studies.

Multidimensional simulations of electron-positron and electron-ion plasmas have indicated that the efficiency of precursor wave emission is roughly an order of magnitude lower than what 1D simulation results suggested \citep[e.g.,][]{iwamotoPersistencePrecursorWaves2017,iwamotoPrecursorWaveAmplification2019}. Incorporating additional dimensional effects would modify Equations (\ref{eq:condition_of_acceleration1}) and (\ref{eq:condition_of_acceleration2}) by a factor of a few. Consequently, we may need to reassess the applicability of the present mechanisms to astrophysical objects accordingly. This issue will also be explored in the near future by large-scale two-dimensional PIC simulations.

\begin{acknowledgments}
This research was supported by JSPS KAKENHI grant No. 23K03407, 
and was supported by MEXT as "Program for Promoting Researches on the Supercomputer Fugaku" (Structure and Evolution of the Universe Unraveled by Fusion of Simulation and AI; Grant Number JPMXP1020230406) and used computational resources of supercomputer Fugaku provided by the RIKEN Center for Computational Science (Project ID:hp250226)
\end{acknowledgments}

\bibliography{ref}
\bibliographystyle{aasjournal}

\end{document}